\documentclass[12pt]{article}
\usepackage{epsfig}
\usepackage{graphicx}

\def\beq{\begin{equation}}
\def\eeq#1{\label{#1}\end{equation}}
\def\eeqn{\end{equation}}

\def\beqa{\begin{eqnarray}}
\def\eeqa#1{\label{#1}\end{eqnarray}}
\def\eeqan{\end{eqnarray}}

\let\bar=\overbar

\def\Dslash{\not{\hbox{\kern-4pt $D$}}}
\def\dslash{\not{\hbox{\kern-2pt $\del$}}}

\def\msb{{\bar{\ssstyle M \kern -1pt S}}}

\usepackage{fancyhdr,graphicx}
\usepackage{graphicx}
\usepackage{float} 
\usepackage{morefloats} 
\usepackage{amssymb} 
\usepackage[square,sort,comma,numbers]{natbib}
\usepackage{subfigure}
\usepackage{multirow}
\usepackage{threeparttable}
\usepackage[table]{xcolor}
\newcommand{\xte}{\hbox{XTE\,J1701$-$462}}
\newcommand{\mxb}{\hbox{MXB\,1659$-$29}}
\newcommand{\ks}{\hbox{KS\,1731$-$260}}
\newcommand{\exo}{\hbox{EXO\,0748$-$676}}

\newcommand{\ig}{\hbox{IGR\,J17480$-$2446}}

\def\Title#1{\begin{center} {\Large {\bf #1} } \end{center}}

\begin{document}

\Title{Inferring neutron stars crust properties from quiescent thermal emission}

\bigskip

\begin{raggedright}

{\it 
Deborah N. Aguilera$^{1,2}$~and Anabela Turlione$^{2}$\\
\bigskip
$^{1}$German Aerospace Center, 
Institute for Space Systems, 
Robert-Hooke Str. 7, 28359~Bremen, 
Germany\\
\bigskip
$^{2}$ Laboratorio Tandar, Comisi\'on Nacional de Energ\'ia At\'omica, 
Av. Gral Paz 1499, 1630~Buenos Aires, 
Argentina\\
{\tt Email: deborah.aguilera@dlr.de}
}

\end{raggedright}

\vspace{0.5cm}

\paragraph{Abstract} The observation of thermal emission from isolated neutron stars
 and the modeling of the corresponding cooling curves has
been very useful to get information on the properties of matter at
very high densities. More recently, the detection of quiescent
thermal emission from neutron stars in low mass X-ray binary systems
 after active periods opened a new window to the physics
of matter at lower densities. Here we analyze a few
sources that have been recently monitored and we show how the
models can be used to establish constraints on the crust
composition and their transport properties, depending on the
astrophysical scenarios assumed.

\section{Introduction}

Neutron stars in binary systems with a low-mass companion star (LMXBs) are most of the time in a quiescent state where almost no accretion occurs and a relative low X-ray luminosity ($< 10^{34}$\,erg\,s$^{-1}$) is observed. But, occasionally, an accretion episode occurs leading to an increase of the observed luminosity of 2-5 orders of magnitude. The accreted material, rich in light elements (H, He), is transferred at rates comparable to Eddington's accretion rates (typically 1-100\%). Once that material reaches the neutron star it undergoes thermal fusion releasing an energy of a few MeV per accreted nucleon, what produces energetic type I X-rays in weekly magnetized systems \cite{Bildsten1997}.
%,Schatz1999}.
When the active phase ends, the X-ray emission decreases dramatically returning to the quiescent levels it had before outbursts. In  quasi-persistent sources (a few cases detected in the last 15 yrs) the accretion periods  last  for a relatively long time: years or decades followed by a reduction in the X-ray luminosity that lasts for about  $10^3$\,days 
after the end of the outburst. These sources are 
\mxb\ \cite{Wijnands2003}, %,Cackett2008}, 
\ks\ \cite{Wijnands2001}, %,Cackett2010b},  
\exo\ \cite{Wolff2008}, %Degenaar2011b, DiazTrigo2011, Degenaar2014nja},  
\xte\ \cite{Fridriksson2011}, %Fridriksson2010}, 
and
 \ig\ \cite{Degenaar2011a}.

\section{Deep crustal cooling and beyond}

\begin{figure}[htb]
\begin{center}
\includegraphics[width=0.55\textwidth, angle=0]{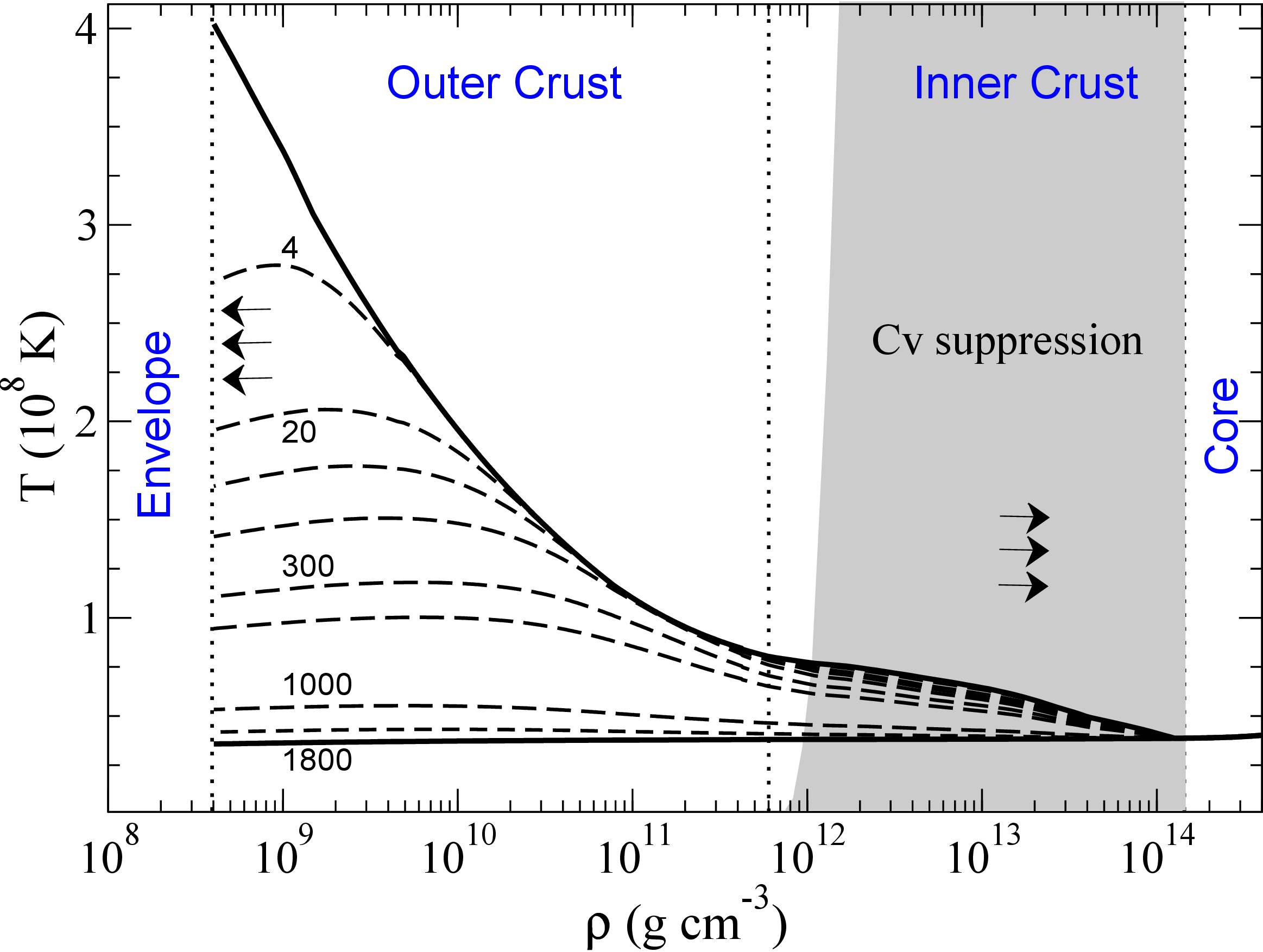}
\caption{Evolution of the initial thermal profile for \mxb. The solid line indicates the temperature across the neutron star crust at $t=0$; 
the dashed-lines show the further evolution of the temperature at $t=4,20, 300, 1000$~days. At $1800$~days the equilibrium 
temperature is reached in the whole crust (solid line at the bottom). The arrows indicate the heat flux direction inside the crust \cite{TurlioneThesis}. 
 }
 \end{center}
\label{fig:22}
\end{figure}

 The quiescent X-ray emission in quasi-persistent sources is thought to be originated in the thermal relaxation of the neutron star crust because the accretion phase  is about as long as the crustal 
diffusion timescale ($\sim$~yr).  Numerical simulations of the crust thermal evolution have been performed recently in \cite{Turlione2015} and previously in \cite{BC2009}. 
Before the active phase the neutron star is supposed to be old enough to have an isothermal interior and its surface 
temperature reflects the core temperature. During the active phase the crust is 
heated up beyond thermal equilibrium by the accretion of matter that compresses the crust and triggers
nuclear reactions.  Once accretion stops the outer layers return to an equilibrium with the interior, a process in which the neutron star
cools down by thermal radiation  from the surface as X-rays,  by heat conduction toward the core and consequent neutrino emission from the core \cite{Brown1998}.
We investigated these processes performing  detailed numerical simulations of the heat flow inside a realistic neutron star crust. In the Fig.~\ref{fig:22} we show the evolution of the initial thermal profile used to study the cooling of \mxb, one of the sources considered as {\it crustal coolers} in \cite{Turlione2015}. 

%%%%%%%%%%%%%%%%%%%%%%%%%%%%%%%%%%%%%%%%%%%%%%%%%%%%%%%%%%%%%%%%%%%%%%%%%
%%%%%%%%%%%%%%%%%%%%%%%%%%%%%%%%%%%%%%%%%%%%%%%%%%%%%%%%%%%%%%%%%%%%%%%%%%%

As a result of this long-term accretion phase, 
the cooling is modified not only by the energy released in the envelope 
(at densities $10^4$--$10^7$g\,cm$^{-3}$) by thermonuclear reactions, 
but also by the energy generated in the inner crust (at $10^{11}$--$10^{13}$g\,cm$^{-3}$) 
by electron captures, neutron emission, 
and density-driven nuclear fusion reactions (pycnonuclear reactions). 
The so-called  {\it deep crustal heating} controls the evolution in the quiescence phase and the comparison of observational data with theoretical models 
allows the investigation of crust properties and ultra-dense matter processes.

\begin{figure}[!htb]
  \vspace{-0.4cm}
\begin{center}
\includegraphics[width=0.6\textwidth, angle=-90]{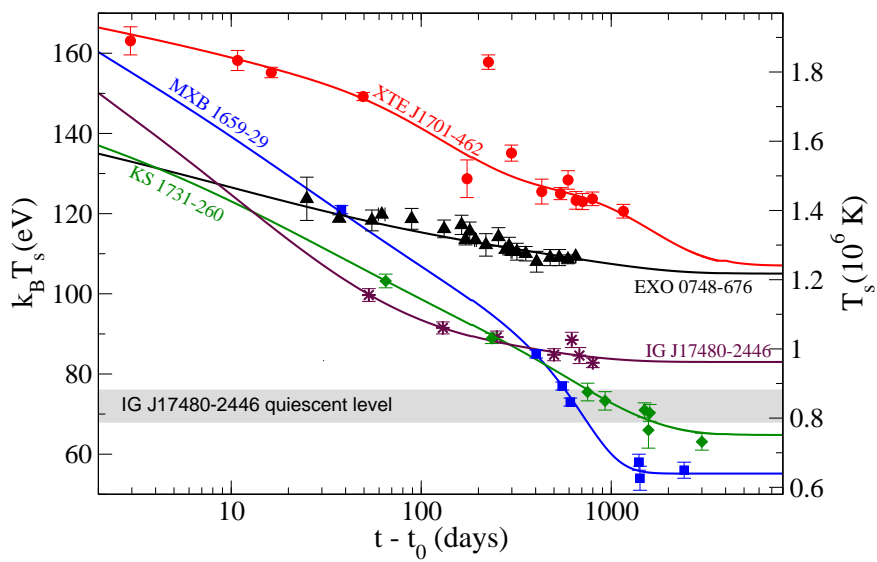}
\caption{Cooling curves obtained from simulations of the  crust thermal evolution (taken from \cite{Turlione2015}). The fits correspond to
       crustal cooling for \mxb\ and \exo, and with addition of: a small energy gap for neutron superfluidity for \ks, 
         shallow heat sources in the outer crust for \xte\ and  a high temperature for the neutron star core for \ig.}
         \vspace{-0.3cm}
\label{fig:24}
\end{center}
\end{figure}

Simulations of the crust relaxation after outbursts for \ks\ and \mxb\  suggested a
rather high thermal conductivity in the outer crust \cite{BC2009} and we found that the evolution of \mxb, \ks\ and \exo\ can be well 
      described within a deep crustal cooling scenario (Fig.~\ref{fig:24}).
%%%%%%%%%%%%%%%%%%%%%%%%%%%%%%%%%%%%%%%%%%%%%%%%%%%%%%%%%%%%%%%%%%%%%%%%%
%%%%%%%%%%%%%%%%%%%%%%%%%%%%%%%%%%%%%%%%%%%%%%%%%%%%%%%%%%%%%%%%%%%%%%%%%%%
 Nevertheless, 
some issues remain open in the search of a model that can explain all the observations consistently, specially for the peculiar emission of \xte\ and \ig\ \cite{PageReddy13, Turlione2015}. 
We proposed different astrophysical scenarios {\it beyond crustal cooling} to explain the variability in \xte\ like residual accretion during quiescence, 
   additional heat sources in the outer crust,  and/or thermal isolation of 
   the inner crust due to a buried magnetic field. 
   We also explained the recent reported temperature of \ig\ with
   an additional heat deposition in the outer crust from shallow sources (Fig.~\ref{fig:24}). 
   Future monitoring of these sources will
determine which scenario is favored, or at least if some of them can be ruled out.

\subsection*{Acknowledgement}

DNA thanks the organizers of the CSQCD IV conference for the fruitful meeting and the support 
from the COST Action MP1304 "Exploring fundamental physics with compact stars".

\end{document}